\documentstyle[12pt]{article}
\newcommand{\sect}[1]{\setcounter{equation}{0}\section{#1}}

\newcommand{\be}{\begin{equation}}
\newcommand{\ee}{\end{equation}}

\def\bkC{{\rm \kern.24em 
\vrule width.05em 
\vrule height1.4ex 
depth-.05ex 
\kern-.26em C}}

\begin{document}

\setlength{\oddsidemargin}{0cm} \setlength{\baselineskip}{7mm}

\begin{normalsize}\begin{flushright}
\end{flushright}\end{normalsize}

\begin{center}
  
\vspace{25pt}
  
{\Large \bf Anisotropic simplicial minisuperspace model in the presence of a
scalar field }

\vspace{40pt}
  
{\sl Crist\'{o}v\~{a}o Correia da Silva}
$^{}$\footnote{e-mail address :
clbc2@damtp.cam.ac.uk}
and {\sl Ruth M. Williams}
$^{}$\footnote{e-mail address : rmw7@damtp.cam.ac.uk}
\\

\vspace{20pt}

DAMTP, Silver Street, Cambridge, \\
CB3 9EW, England \\

\end{center}

\vspace{40pt}

\begin{center} {\bf ABSTRACT } \end{center}
\vspace{12pt}
\noindent

We study an anisotropic simplicial minisuperspace model with a
cosmological constant and a massive scalar field. After obtaining the
classical solutions we  then compute the semiclassical approximation
of the no-boundary wave function of the Universe along a steepest
descents contour passing through classical Lorentzian solutions. The
oscillatory behaviour of the resulting wave funtion is consistent with
the  prediction of classical Lorentzian spacetime
for the late Universe.

\vspace{24pt}

\vfill

\newpage

\sect{Introduction}

The Euclidean path integral formulation of quantum gravity is arguably
the most promising of all non-stringy approaches to quantum
cosmology. The canonical quantisation approach translated by the
Wheeler-DeWitt equation does not possess the flexibility of the path
integral formulation as it cannot account for topology changing
processes. The Euclidean path integral formulation with its sum over
geometries naturally allows these processes to occur as it considers a
sum over different topologies. As important as how to obtain the
dynamics of the quantum state of the Universe is to use the correct
prescription for the boundary condition needed to obtain the 
wave function of the Universe. The most natural boundary condition, the
Hartle-Hawking no boundary proposal,\cite{hawk},  requires that the sum over
geometries be restricted to those four-geometries which have no
boundary except for the prescribed three-geometry where the arguments
of the wave function are defined. This is specially well adapted to the
path integral formalism as it can be easily implemented in a sum over topologies context.  
However, even this proposal is plagued by at least two main
problems. Firstly, the Euclidean gravitational action is not bounded
from below which leads to the divergence of the Euclidean path
integral. Secondly,  there is no clear prescription for the
correct integration contour to use. In  \cite{h2}, Hartle proposes
the use of the steepest descents contour in the space of complex
metrics as the solution to both problems. Furthermore 
by choosing  the steepest descents contour passing through the classical
solutions of the theory, he  made it very likely that the path
integral be dominated by classical four-geometries, i.e., solutions of
Einstein's equations and stationary points of the path integral, as
desired for any wave function that is intended to represent our current Universe. 
Note that in this view, the fact that an integration solely over
real-valued Euclidean geometries does not yield a convergent result
for the path integral, is actually a good thing, for such a  path integral would never predict the oscillatory behaviour in the late Universe that  traditionally represents classical Lorentzian space-time.

As it is clearly impossible to calculate the full wave function
integrated over all possible metric degrees of freedom, it is usual to
explore approximated models in which the infinite degrees of freedom
are  reduced to only a few. The loss of generality induced by this
reduction is not as bad it might seem at first,  if we remember the
relative symmetry  of the observed Universe. These models are
generically called minisuperspace models, for they "exist" in small,
few-dimensional, subspaces contained in  the infinite-dimensional
superspace which is the space of all three-metrics. Particularly
useful among minisuperspace models are the the ones based on Regge
calculus. Such simplicial minisuperspace models were introduced by
Hartle \cite{h1}. In such models one typically takes the simplicial
complex which models the topology of interest to be fixed and the
square edge length assignments play the role of the metric degrees of
freedom. The summation over edge lengths models the continuum
integration over the metric tensor. This approach has several
advantages. First by treating the four-geometry directly it is more
adequate to deal with the Hartle-Hawking proposal, \cite{hawk}, (with its four-dimensional nature), than the usual $3+1$ ADM decomposition of space-time, where a careful study of how the four-geometry closes off at the beginning of the universe is essential. 
Second, by discretizing space-time the classical equations become algebraic which makes it easier to find classical solutions which are essential to the semiclassical approximation. Third, it offers the possibility of systematic improvement.

In \cite{h2}, Hartle studied the simplest simplicial minisuperspace
model, the $\alpha _{4}$ triangulation of the three-sphere. He assumed
isotropy which leads to all  boundary edge lengths being  equal. Later
in \cite{birm}, Birmingham generalised his work to isotropic
triangulations of Lens spaces $L(p,1)$. Another model was considered
by Louko and Tuckey in \cite{louko} working in a minisuperspace anisotropic
$2+1$-dimensional cosmological model.  In  \cite{fur}, Furihata considered a more general anisotropic triangulation of the three-sphere. Finally, in \cite{meu}, we generalised the isotropic models to incorporate all simplicial four-geometries that are cones over closed connected simplicial three-manifolds, i.e., simplicial four-conifolds, and for the first time introduced a massive scalar field. In all these cases it was possible to find a steepest descent contour and to prove that the resulting wave function shows oscillatory behaviour for large spatial geometries. We now propose  to study an extension of Furihata's model to incorporate a matter sector. We investigate the existence of a steepest descent contour and the properties of the resulting wave function.

\section{General Regge Formalism}

A convenient way of defining an $n-$simplex is to specify the
coordinates of its $(n+1)$ vertices, $\sigma =[0,1,2,...,n]$. By
specifying the squared values of the lengths of the edges $[i,j]$,
$s_{ij}$, we fix the simplicial metric on the simplex.

\be
\label{simpmet}
g_{ij}(s_{k})=\frac{s_{0i}+s_{0j}-s_{ij}}{2}
\ee
where $i,j=1,2,..n$.

So if we triangulate a smooth manifold $M$ endowed with a metric $g_{\mu \nu}$ by a homeomorphic simplicial manifold 
${\cal{M}}$, the metric information is transferred to the simplicial metric of that simplicial complex

\be 
g_{\mu \nu}(x) \longrightarrow   g_{ij}(\{s_{k}\})=\frac{s_{0i}+s_{0j}-s_{ij}}{2}
\ee

In the continuum framework the sum over metrics is implemented through a functional integral over the metric components $\{g_{\mu \nu}(x)\}$. In the simplicial framework the metric degrees of freedom are the squared edge lengths, and so the  functional integral is replaced by a simple multiple integral over the values of the edge lengths. But not all edge lengths have equal standing. Only the ones associated with the interior of the simplicial complex get to be integrated over. The boundary edge lengths remain after the sum over metrics and become the arguments of the wavefunction of the universe.

\be
\int Dg_{\mu \nu }(x) \longrightarrow \int D\{s_{i}\}=\prod \int d\mu (s_{i}) 
\ee

In the simplicial framework the fact that the geometry of the complexes is completely fixed by the specification of the squared values of all edge lengths, means that all geometrical quantities, such as volumes and curvatures, can be expressed completely in terms of those edge lengths. Consequently the Regge action (the simplicial analogue of the Einstein action for GR) associated with a complex of known topology can be expressed exclusively in terms of those edge lengths.

The Euclideanized Einstein action for a smooth $4-$manifold $M$ with boundary $\partial M$, and endowed with a $4-$metric, $g_{\mu \nu }$, and a scalar field $\Phi $ with mass $m$, is

\begin{eqnarray*}
I[M,g_{\mu \nu},\phi ]&=&-\int _{M}d^{4}x\sqrt{g}\frac{(R-2\Lambda
                       )}{16\pi G}-\int _{\partial
                       M}d^{3}x\sqrt{h}\frac{K}{8\pi G}+ \\
                       &+& \mbox{} \frac{1}{2}\int  _{M}d^{4}x\sqrt{g}(\partial _{\mu}\phi \partial ^{\mu} \phi +m^{2} \phi ^{2})   
\end{eqnarray*}
where $K$ is the extrinsic curvature.

Its simplicial analogue will be the Regge action for a combinatorial
$4-$manifold, ${\cal {M}}$, with squared edge lengths $\{s_{k}\}$, and
with a scalar field taking values $\{\phi _{v}\}$ for each vertex $v$
of ${\cal {M}}$, \cite{ruth3}:

\begin{eqnarray*}
 I[{\cal {M}},\{s_{k}\},\{\phi _{v}\}]&=&\frac{-2}{16\pi G}\sum _{\sigma _{2}^{i}} V_{2}(\sigma _{2}^{i})\theta (\sigma _{2}^{i}) +  {\frac{2\Lambda }{16\pi G}} \sum _{\sigma _{4}}V_{4}(\sigma _{4}) \\
                                      &-&{\frac{2}{16\pi G}}\sum _{\sigma _{2}^{b}} V_{2}(\sigma _{2}^{b})\psi (\sigma _{2}^{b})+\frac{1}{2}\sum _{\sigma _{1}=[ij]} \tilde{V}_{4}(\sigma _{1})\frac{(\phi _{i}-\phi _{j})^{2}}{s_{ij}}\\
                                      &+& \mbox{}\frac{1}{2}\sum _{j}\tilde{V}_{4}(j)m^{2}\phi _{j}^{2}   
\end{eqnarray*}
where:

\begin{itemize}

 \item $\sigma _{k}$ denotes a $k-$simplex belonging to the set $\Sigma _{k}$ of all  $k-$simplices in  ${\cal {M}}$.

\item $\theta(\sigma _{2}^{i})$, is the deficit angle associated with the interior $2-$simplex $\sigma _{2}^{i}=[ijk]$

\be
  \theta(\sigma _{2}^{i})=2\pi -\sum _{\sigma _{4}\in St(\sigma _{2}^{i})}\theta _{d}(\sigma _{2}^{i},\sigma _{4})   
\ee

and $\theta _{d}(\sigma _{2}^{i},\sigma _{4})$ is the dihedral angle between the $3-$simplices $\sigma _{3}=[ijkl]$ and $\sigma ^{'}_{3}=[ijkm]$, of $\sigma _{4}=[ijklm]$ that intersect at  $\sigma _{2}^{i}$. Its full expression is given by \cite{birm}.

\item   $\psi (\sigma _{2}^{b})$ is the deficit angle associated with the boundary $2-$simplex $\sigma _{2}^{b}$:

\be
  \psi (\sigma _{2}^{b})=\pi -\sum _{\sigma _{4}\in St(\sigma _{2}^{b})}\theta _{d}(\sigma _{2}^{b},\sigma _{4})    
\ee

\item $V_{k}(\sigma _{k})$ for $k=2,3,4$ is the $k-$volume associated with the $k-$simplex, $\sigma _{k}$, and once again their explicit expressions in terms of the squared edge lengths are given by  \cite{birm}.

\item  $\tilde{V}_{4}(\sigma _{1})$ is the $4-$volume in the simplicial complex ${\cal{M}}$, associated with the edge $\sigma _{1}$, i.e., the volume of the space occupied by all points of ${\cal{M}}$ that are closer to $\sigma _{1}$ than to any other edge of  ${\cal{M}}$. The same holds for $\tilde V_{4}(j)$ where $j$ represents all vertices of ${\cal{M}}$.

\end{itemize}

It is easy to see that both $\tilde{V}_{4}(\sigma _{1})$ and $\tilde
V_{4}(j)$, can be expressed exclusively in terms of the edge lengths
$\{s_{k}\}$. All  these expressions remain valid if we consider smooth conifolds and their combinatorial counterparts.

If we are working in a simplicial minisuperspace model based on a
simplicial four-manifold ${\cal M}_{4}$, with boundary edges
${s_{b}}$, interior edges ${s_{i}}$, and in the presence of a scalar
field taking values, ${\phi_{k}}$, at the vertices $k\in {\cal M}_{4} $, then the wave function of the Universe is of the type:

\be
\label{2}
\Psi [\partial {\cal{M}},\{s_{b}\},\{\phi _{b}\}]=\int D\{s_{i}\}D\{\phi _{i}\}e^{-I[{\cal{M}}^{4},\{s_{i}\},\{s_{b}\},\{\phi _{i}\},\{\phi _{b}\}]}  
\ee
where

\begin{itemize}

\item $ \{s_{i}\}$ are the squared lengths of the interior edges 

\item  $\{s_{b}\}$ are the squared lengths of the boundary  edges 

\item  $\{\phi_{i}\}$ are the values of the field at the interior
vertices    

\item  $\{\phi_{b}\}$ are the values of the field at the boundary vertices

\end{itemize}

Although the functional integral over metrics has been written explicitly
 in terms of the edge lengths, this expression is still
heuristic because we still need to specify  the
measure,  and the integration contour to be used.

\sect{The Model}

As in \cite{fur} our simplicial complex can be seen as a cone, 
${\cal{M}}^{4}=0*{\cal M}^{3}$, over an anisotropic  triangulation of
the three-sphere, ${\cal M}^{3}=\alpha_{4}(a,b)$. See figure 1. There are
$10$ boundary edges, five  with squared length, $a$, and
the other five with squared length $b$. There are also five interior
edges connecting the five boundary vertices ${1,2,3,4,5}$ to the
interior vertex $0$, and they all have squared length $s_{i}$. Note
that although the model is anisotropic because there are two kinds of
boundary edges, it is still homogeneous because all boundary vertices
are connected to  the same number and kind of edges. Furthermore, we
consider the scalar field to take the same value, $\phi_{b}$ at all
boundary vertices, and $\phi_{i}$ at the interior vertex $0$.

For simplicity we will use new variables,

\be
\xi =\frac{s_{i}}{a}
\ee
\be
 S=\frac{H^{2}a}{l^{2}}
\ee
\be
 \beta =\frac{b}{a}
\ee

where $H^{2}=l^{2}\Lambda /3$, and $l^{2}=16\pi G$ is the Planck length. We shall work in units where $c=\hbar =1$.

The Euclidean action $I$, is then a function of $\xi, S, \beta, \phi_{b},\phi_{i}$, and the minisuperspace wave function is

\be
\label{ppsi}
\Psi [  S,\beta,\phi _{b}]=\int_{C} D\xi  D\phi _{i}e^{-I[\xi, S,\beta ,\phi _{i},\phi _{b}]}   
\ee

To calculate the explicit expression of the action we must compute the volumes and deficit angles associated with the various $p-simplices$ in ${\cal M}_{4}$.

There is only one kind of $4-simplex$, eg., $[01234]$, and its volume is

\be
V_{4}[01234]=\frac{1}{4!}\biggl(\frac{a}{2}\biggr)^{2}\sqrt{-\beta ^{2}+3\beta -1}\sqrt{4\xi(\beta+1)-1-\beta -\beta^{2}}
\ee

There is also only one kind of boundary tetrahedron, eg., $[1234]$, and its volume is

\be
V_{3}^{i}[1234]=\sqrt{\frac{2}{(3!)^{2}}\biggl(\frac{a}{2}\biggr)^{3}(\beta +1)(-\beta^{2}+3\beta-1)}
\ee

The existence of two different boundary edges means that there are two kinds of interior tetrahedra

\be
V_{3A}^{b}[0123]=\biggl(\frac{1}{3!}\biggr)\biggl(\frac{a}{2}\biggr)^{3/2}\sqrt{2\beta\{(4-\beta)\xi-1\}} 
\ee

\be
V_{3B}^{b}[0124]=\biggl(\frac{1}{3!}\biggr)\biggl(\frac{a}{2}\biggr)^{3/2}\sqrt{2\{(4\beta-1)\xi-\beta^{2}\}} 
\ee
There are also two kinds of interior triangles, whose areas are

\be
V_{2A}^{i}[012]=\frac{a}{2}\sqrt{\xi-1/4}
\ee

\be
V_{2B}^{i}[013]=\frac{a\sqrt{\beta}}{2}\biggl(\xi-\frac{\beta}{4}\biggr)^{1/2}
\ee

Finally there are also two kinds of boundary triangles 

\be
V_{2A}^{b}[124]=\frac{a}{2}\sqrt{\beta-1/4}
\ee

\be
V_{2B}^{b}[123]=\frac{a\sqrt{\beta}}{2}\biggl(1-\frac{\beta}{4}\biggr)^{1/2}
\ee

The $4-volumes$ associated with each boundary vertex $b$, and with the interior vertex $0$ are

\be
\tilde{V}_{4}[b]=\frac{4}{5}V_{4}[01234],\:\:\:\;\; \tilde{V}_{4}[0]=V_{4}[01234]
\ee

The $4-volume$ associated with each interior edge $[0,b]$ is

\be
\tilde{V}_{4}[0,b]=\frac{2}{5}V_{4}[01234]
\ee

The deficit angles associated with each kind of triangle are

\be
\theta [012]=2\pi-\arccos{h_{1}}-2\arccos{h_{2}}
\ee
\be
\theta [013]=2\pi-\arccos{h_{3}}-2\arccos{h_{4}}
\ee
\be
\psi [123]=\pi-2\arccos{h_{5}}
\ee
\be
\psi [124]=\pi-2\arccos{h_{6}}
\ee

where the dihedral angles are

\be
h_{1}=\frac{2(\beta^{2}-2\beta+2)\xi-\beta^{2}+\beta-1}{2\beta(4-\beta)(\xi-\xi_{3A})}
\ee

\be
h_{2}=\frac{2(3\beta-2)\xi-\beta^{2}-\beta+1}{2\sqrt{\beta(4\beta-1)(4-\beta)}\sqrt{(\xi-\xi_{3A})(\xi-\xi_{3B})}}
\ee

\be
h_{3}=\frac{2(2\beta^{2}-2\beta+1)\xi-\beta(\beta^{2}-\beta+1)}{2(4\beta-1)(\xi-\xi_{3B})}
\ee

\be
h_{4}=\sqrt{\frac{\beta}{(4-\beta)(4\beta-1)}}\frac{2(3-2\beta)\xi+\beta^{2}-\beta-1}{2\sqrt{(\xi-\xi_{3A})(\xi-\xi_{3B})}}
\ee

\be
h_{5}=\frac{1}{2}\sqrt{\frac{-\beta^{2}+3\beta-1}{(4\beta-1)(\beta+1)}}\frac{1}{\sqrt{\xi-\xi_{3B}}}
\ee

\be
h_{6}=\frac{1}{2}\sqrt{\frac{\beta(-\beta^{2}+3\beta-1)}{(4-\beta)(\beta+1)}}\frac{1}{\sqrt{\xi-\xi_{3A}}}
\ee
where

\be
\xi_{3A}=\frac{1}{4-\beta}\:\;\;\:\:\:\:\:\:\:\xi_{3B}=\frac{\beta^{2}}{4\beta-1}
\ee
are the values of $\xi$ for which  the interior
tetrahedra become degenerate.

The Regge minisuperspace action then becomes

\begin{eqnarray*}
I[\xi ,S,\beta ,\phi _{i},\phi _{b}]&=&-\biggl(\frac{S}{H^{2}}\biggr)\biggl\{
\frac{5}{2}\sqrt{4\beta-\beta^{2}}(\pi-2\arccos{h_{6}})+\frac{5}{2}\sqrt{4\beta-1}(\pi-2\arccos{h_{5}})\\
                             &+& 5\sqrt{\xi-1/4}(2\pi-\arccos{h_{1}}-2\arccos{h_{2}})\\
                             &+&5\sqrt{\beta}\sqrt{\xi-\frac{\beta}{4}}(2\pi-\arccos{h_{3}}-2\arccos{h_{4}})\\
                             &-&\frac{1}{48}\sqrt{\beta+1}\sqrt{-\beta^{2}+3\beta-1}\frac{\sqrt{\xi-\xi_{4}}}{\xi}(\phi_{i}-\phi_{b})^{2}l^{2}\biggr\}\\
                             &+&\mbox{}\biggl(\frac{S}{H}\biggr)^{2}\biggl(\frac{5}{8}\biggr)\sqrt{\beta+1}\sqrt{-\beta^{2}+3\beta-1}\sqrt{\xi-\xi_{4}}\biggl[1+\frac{1}{60}\biggl(\frac{ml}{H}\biggr)^{2}(4\phi_{b}^{2}l^{2}+\phi_{i}^{2}l^{2})\biggr]
\end{eqnarray*}
\be
\label{actionI}
\ee

where 
\be
\xi_{4}=\frac{\beta^{2}+\beta+1}{4\beta+4}
\ee
is the the value of $\xi$ for which  the $4-simplices$ become
degenerate.

Since the wave function is to be obtained as an integral over $\xi$
and $\phi_{i}$, it is essential we study, term by term, the analytic and asymptotic
properties of the action as a function of these variables. It is easy
to see that the action is an analytic function for all values of
$\phi_{i}$. However, $I$ as a function of $\xi$ has several square
root branch points, and logarithmic branching points and infinities.

Note that a term $\arccos{u(z)}$ has branch points at $u(z)=+1, -1$,
and at $u(z)=\infty$. The associated branch cuts are usually taken
to be $(-\infty ,-1]\cup [1,+\infty )$ . Since

$$\arccos{u(z)}=-i\log \biggl(u(z)+\sqrt{u(z)^{2}-1}\biggr)$$

then we see that there are logarithmic singularities when
$u(z)=\infty$. The table below  shows the logarithmic branching points and
infinities associated with the dihedral angles. Furthermore, associated
with the square root terms we have branch points at  $\xi=\frac{1}{4}$,
 $\xi=\frac{\beta}{4}$, and $\xi=\xi_{4}$.

\vspace{55pt}

\begin{tabular}{|c|c|c|c|}  \hline 
\hspace{.1in}Dihedral angles\hspace{.4in}&\hspace{.4in}$-1$\hspace{.4in}
  &\hspace{.4in}$+1$\hspace{.4in}
  &\hspace{.4in}$\infty $\hspace{.4in} \\ \hline
\hspace{.4in}$h_{1}(\xi)$\hspace{.4in}   & \hspace{.4in}$\xi_{4}$ \hspace{.4in}      & \hspace{.4in}  $1/4$\hspace{.4in}  &\hspace{.4in}$\xi_{3A}$\hspace{.4in}      \\ \hline

\hspace{.4in}$h_{2}(\xi)$\hspace{.4in}   & \hspace{.4in}$1/4,
\xi_{4}$\hspace{.4in}      & \hspace{.4in} $1/4, \xi_{4}$\hspace{.4in} &\hspace{.4in}$\xi_{3A},\xi_{3B}$\hspace{.4in}        \\ \hline

\hspace{.4in}$h_{3}(\xi)$\hspace{.4in} & \hspace{.4in}$\xi_{4}$\hspace{.4in}      & \hspace{.4in} $\beta/4$\hspace{.4in} &\hspace{.4in}$\xi_{3B}$\hspace{.4in}        \\ \hline
\hspace{.4in}$h_{4}(\xi)$\hspace{.4in} &
\hspace{.4in}$\beta/4, \xi_{4}$\hspace{.4in}      &
\hspace{.4in} $\beta/4, \xi_{4}$\hspace{.4in} &\hspace{.4in}$\xi_{3A},\xi_{3B}$\hspace{.4in}        \\ \hline
\hspace{.4in}$h_{5}(\xi)$\hspace{.4in}  &
\hspace{.4in}$\xi_{4}$\hspace{.4in}      & \hspace{.4in} $\xi_{4}$\hspace{.4in}  &\hspace{.4in}$\xi_{3B}$\hspace{.4in}       \\ \hline
\hspace{.4in}$h_{6}(\xi)$\hspace{.4in}  & \hspace{.4in}$\xi_{4}$\hspace{.4in}      & \hspace{.4in} $\xi_{4}$\hspace{.4in}  &\hspace{.4in}$\xi_{3A}$\hspace{.4in}       \\ \hline

\end{tabular}

\vspace{45pt}

Implementing the necessary branch cuts it is easy to conclude that the
Riemann surface of the action is composed of an infinite number of
sheets, each with a branch cut $(-\infty ,\xi_{4}]$. However, as in 
\cite{meu}, only three sheets will be relevant for the computation of
the path integral. We define the first sheet ${\bkC} _{1}$ of $I[\xi ]$ as the sheet where the terms in $\arccos (z)$ assume their principal  values. So the action in the first sheet will be formally equal to the original expression $(\ref{actionI} )$. Note that with the first sheet defined in this way, for real $\xi >\xi_{4}$ the volumes and deficit angles are all real, leading to a real Euclidean  action for $\xi \in [\xi_{4},+\infty )$ on the first sheet. 
On the other hand, when $\xi $ is real and less than $min(1/4,\beta/4)$ in the first sheet , the volumes become pure imaginary and the Euclidean action becomes pure imaginary. For all other points of this first sheet the action is fully complex.

Note that since we are only interested in geometries in which the
boundary three-metric is positive definite, we must require that the
volume of the boundary three-simplices be positive, which is
equivalent to requiring $\frac{3-\sqrt{5}}{2}<\beta <
\frac{3+\sqrt{5}}{2}$. Furthermore, since the simplicial metric in each
four-simplex is real if and only if $\xi$ is real, then the  simplicial
geometries built out of these $4-$simplices will be real only when
$\xi $ is real. Finally, computing the eigenvalues
$(\lambda_{1}, \lambda_{2}, \lambda_{3}, \lambda_{4})$ of the simplicial
metric,  $g_{ij}$, we obtain

\be
\lambda_{1}=\frac{1}{4}\biggl[8\xi-(\beta+1)-\sqrt{[8\xi-2(\beta+1)]^{2}+(\beta-1)^{2}}\biggr] 
\ee
\be
\lambda_{2}=\frac{1}{4}\biggl[8\xi-(\beta+1)+\sqrt{[8\xi-2(\beta+1)]^{2}+(\beta-1)^{2}}\biggr] 
\ee
\be
\lambda_{3}=\frac{1}{4}[1+\beta+\sqrt{5}(\beta-1)]
\ee
\be
\lambda_{4}=\frac{1}{4}[1+\beta-\sqrt{5}(\beta-1)]
\ee

  It is easy
to see that since $\frac{3-\sqrt{5}}{2}<\beta <\frac{3+\sqrt{5}}{2}$,
then $\lambda_{2}, \lambda_{3}, \lambda_{4}$, are all positive and so the
signature  is Euclidean $(++++)$, when $\lambda_{1}>0$, i.e., when $\xi>\xi_{4}$
and Lorentzian $(-+++)$, when $\lambda_{1}<0$, i.e., when $\xi<\xi_{4}$. So we see that for real
$\xi >\xi_{4}$ we have real Euclidean signature geometries, with real
Euclidean action, and for real $\xi < min(\frac{1}{4},\frac{\beta}{4})$, we have real Lorentzian signature geometries with pure imaginary Euclidean action.

When the action is continued in $\xi $ once around all finite branch
points $\xi = 1/4,\beta /4, \xi_{3A},\xi_{3B}$, and $ \xi_{4}$, we reach what shall be called the second sheet . It is easy to conclude that the action in this second sheet is just the negative of the action in the first sheet.

\be
I^{I}[\xi,\beta ,S,\phi _{i},\phi _{b}]=-I^{II}[\xi,\beta ,S,\phi _{i},\phi _{b}] 
\ee

Once in the second sheet, if we encircle the branch points in such a
way that we cross the branch cut, $(-\infty ,\xi_{4}]$, between $1/4$
and $\beta/4$, we arrive at what we shall call the third sheet. By
doing this the terms $\arccos{h_{1}}$, and $\arccos{h_{2}}$, change
signs, but $\sqrt{\xi-1/4}$ does not and so the action in this third
sheet is different from in the first sheet

\begin{eqnarray*}
I_{III}[\xi ,S,\beta ,\phi _{i},\phi _{b}]&=-&\biggl(\frac{S}{H^{2}}\biggr)\biggl\{
\frac{5}{2}\sqrt{4\beta-\beta^{2}}(\pi-2\arccos{h_{6}})+\frac{5}{2}\sqrt{4\beta-1}(\pi-2\arccos{h_{5}})\\
                             &-& 5\sqrt{\xi-1/4}(2\pi+\arccos{h_{1}}+2\arccos{h_{2}})\\
                             &+&5\sqrt{\beta}\sqrt{\xi-\frac{\beta}{4}}(2\pi-\arccos{h_{3}}-2\arccos{h_{4}})\\
                             &-&\frac{1}{48}\sqrt{\beta+1}\sqrt{-\beta^{2}+3\beta-1}\frac{\sqrt{\xi-\xi_{4}}}{\xi}(\phi_{i}-\phi_{b})^{2}l^{2}\biggr\}\\
                             &+&\mbox{}\biggl(\frac{S}{H}\biggr)^{2}\biggl(\frac{5}{8}\biggr)\sqrt{\beta+1}\sqrt{-\beta^{2}+3\beta-1}\sqrt{\xi-\xi_{4}}\biggl[1+\frac{1}{60}\biggl(\frac{ml}{H}\biggr)^{2}(4\phi_{b}^{2}l^{2}+\phi_{i}^{2}l^{2})\biggr]
\end{eqnarray*}
\be
\label{actionIII}
\ee

\subsection{Asymptotic Behaviour of the Action}

In any discussion of the convergence of an integral over an infinite
contour the asymptotic behaviour  of the integrand,  when  $\mid \xi \mid
\rightarrow \infty $, is essential.

In the first sheet when $ \xi \rightarrow \infty $ the action behaves like

\begin{eqnarray*}
I^{I}[\xi,\beta,S,\phi_{i},\phi_{b}]&\sim &\frac{5}{8}\sqrt{(\beta+1)(-\beta^{2}+3\beta-1)}\\
                                &\times &  \mbox{}  \biggl[1+\frac{1}{60}\frac{m^{2}l^{2}}{H^{2}}(4\phi_{b}^{2}l^{2}+\phi_{i}^{2}l^{2})\biggr]\frac{S}{H^{2}}(S-S_{crit}^{I})\sqrt{\xi}
\end{eqnarray*}
\be
\label{Ias}
\ee
where

\begin{eqnarray*}
S_{crit}^{I}&=&\frac{8}{\sqrt{(\beta+1)(-\beta^{2}+3\beta-1)}}\biggl[(2\pi-\arccos{h_{1}^{\infty}}-2\arccos{h_{2}^{\infty}})\\
        &+& \mbox{} \sqrt{\beta}(2\pi-\arccos{h_{3}^{\infty}}-2\arccos{h_{4}^{\infty}})\biggr] \biggl[1+\frac{1}{60}\frac{m^{2}l^{2}}{H^{2}}(4\phi_{b}^{2}l^{2}+\phi_{i}^{2}l^{2})\biggr]^{-1}
\end{eqnarray*}
\be
\label{s1crit}
\ee

and

\be
h_{1}^{\infty}=\frac{\beta^{2}-2\beta+2}{\beta(4-\beta)}
\ee

\be
h_{2}^{\infty}=\frac{3\beta-2}{\sqrt{\beta(4\beta-1)(4-\beta)}}
\ee
\be
h_{3}^{\infty}=\frac{2\beta^{2}-2\beta+1}{4\beta-1}
\ee

\be
h_{4}^{\infty}=\sqrt{\frac{\beta}{(4-\beta)(4\beta-1)}}(3-2\beta)
\ee

For the second sheet the asymptotic behaviour of the action is just the negative
of that in the first sheet. For the third sheet we have

\begin{eqnarray*}
I^{III}[\xi,\beta,S,\phi_{i},\phi_{b}]&\sim &\frac{5}{8}\sqrt{(\beta+1)(-\beta^{2}+3\beta-1)}\\
                                &\times &  \mbox{}  \biggl[1+\frac{1}{60}\frac{m^{2}l^{2}}{H^{2}}(4\phi_{b}^{2}l^{2}+\phi_{i}^{2}l^{2})\biggr]\frac{S}{H^{2}}(S+S_{crit}^{III})\sqrt{\xi}
\end{eqnarray*}
\be
\label{IIIas}
\ee
where

\begin{eqnarray*}
S_{crit}^{III}&=&\frac{8}{\sqrt{(\beta+1)(-\beta^{2}+3\beta-1)}}\biggl[(2\pi+\arccos{h_{1}^{\infty}}+2\arccos{h_{2}^{\infty}})\\
        &-& \mbox{} \sqrt{\beta}(2\pi-\arccos{h_{3}^{\infty}}-2\arccos{h_{4}^{\infty}})\biggr] \biggl[1+\frac{1}{60}\frac{m^{2}l^{2}}{H^{2}}(4\phi_{b}^{2}l^{2}+\phi_{i}^{2}l^{2})\biggr]^{-1}
\end{eqnarray*}
\be
\label{s3crit}
\ee

\section{Classical Solutions}

The classical simplicial geometries are the extrema of the Regge action we  obtained above. In our minisuperspace model there are two degrees of freedom $\xi ,\phi _{i}$. So the Regge equations of motion will be:

\be
\frac{\partial I}{\partial \xi}=0
\label{classic1}
\ee
and

\be
\frac{\partial I}{\partial \phi _{i}}=0
\label{cla2}
\ee

They are to be solved for the values of $\xi ,\phi _{i}$, subject to
the fixed boundary data $S, \beta,\phi _{b}$. The classical solutions will
thus be of the form $\overline{\xi}(S,\beta, \phi _{b})$, and
$\overline{\phi} _{i}(S,\beta,\phi _{b})$. The solution $\overline{\xi}(S,\beta,
\phi _{b})$ completely determines the simplicial geometry.

Note that we shall be working  on the first sheet. Of course, since on
the second sheet the action is just the negative of this,  the
equations of motion are the same. And obviously every classical
solution   $\overline{\xi}_{I}(S,\beta ,\phi _{b})$ located on the
first sheet will have a counterpart $\overline{\xi}_{II}$ of the same
numerical  value, but located on the second sheet, and so with an
action of opposite sign, $I[\overline{\xi}_{I}(S,\beta , \phi _{b})]=-I[\overline{\xi}_{II}(S,\beta ,\phi _{b})]$. So the classical solutions occur in pairs.

Inserting the second equation, $(\ref{cla2})$, 

\be
\label{cleqphi}
\phi _{i}=\frac{\phi _{b}}{1+\frac{1}{2}\frac{m^{2}l^{2}}{H^{2}}\xi S}  
\ee
in the first equation $(\ref{classic1})$,  we obtain a very long cubic equation on S for each value
of $\xi$,
given fixed $\beta$ and $\phi_{b}$. This equation can then be solved numerically
for $\xi$, and by inverting the resulting solutions we obtain
three branches of solutions $\xi=\xi _{cl}(S,\beta,\phi_{b})$. For
obvious physical reasons we shall accept only solutions with real
positive $S$. In figure $2$ we show one such solution for a fixed
value of $\beta $ and $\phi_{b}$.

The general form of this solution is similar to the ones associated
with other values of  $\beta $ and $\phi_{b}$. We see that in general
as $\xi \rightarrow \xi_{0}=min(1/4,\beta/4)$, the value of $S$ diverges to
$+\infty$. It is this branch that will represent the classical
solutions for the late Universe. The fact that these solutions are
Lorentzian means that semiclassically the wave function of the Universe should be    
dominated by the contribution coming from classical Lorentzian
spacetimes like our own, as desired.

As in previous cases the domain of solutions is divided by the line
$S=S^{I}_{crit}(cl)$, where 

\begin{eqnarray*}
S^{I}_{crit}(cl)&=&S^{I}_{crit}(\phi_{i}=\phi_{i}^{cl})=\frac{8}{\sqrt{(\beta+1)(-\beta^{2}+3\beta-1)}}\biggl[(2\pi-\arccos{h_{1}^{\infty}}\\
                &-& \mbox{}2\arccos{h_{2}^{\infty}})+ \sqrt{\beta}(2\pi-\arccos{h_{3}^{\infty}}-2\arccos{h_{4}^{\infty}})\biggr] \biggl(1+\frac{2}{15}K\phi_{b}^{2}\biggr)^{-1}
\end{eqnarray*}
where $K=\frac{m^{2}l^{2}}{2H^{2}}$.

 We see that for $0<S <S_{crit}^{I}(cl)$, we will have:

\begin{itemize}

\item  Two pairs of real Lorentzian signature solutions $\overline{\xi
}_{I}^{L1}(S,\beta,\phi _{b})=\overline{\xi }_{II}^{L1}(S,\beta,\phi _{b}) \in
(-\infty ,\xi_{0}]$, and $\overline{\xi }_{I}^{L2}(S,\beta,\phi _{b})=\overline{\xi }_{II}^{L2}(S,\beta,\phi _{b}) \in (-\infty ,\xi_{0}]$ with pure imaginary Euclidean actions.

\item One pair of real Euclidean signature solutions $\overline{\xi
}_{I}^{E}(S,\beta,\phi _{b})=\overline{\xi }_{II}^{E}(S,\beta,\phi _{b}) \in
[\xi_{4},+\infty )$,  with real Euclidean action.

\end{itemize}

 For $S >S_{crit}^{I}(cl)$ we have:

\begin{itemize}

\item  Only one pair of real solutions $\overline{\xi }_{I}(S,\beta,\phi _{b})=\overline{\xi }_{II}(S,\beta.\phi _{b}) \in (-\infty ,\xi_{0}]$ that correspond to Lorentzian signature simplicial metrics, and whose Euclidean actions, though symmetric, are both pure imaginary. 

\end{itemize}

$$ I[\overline{\xi }_{I}(S,\beta,\phi _{b})]=-I[\overline{\xi }_{II}(S,\beta,\phi _{b})]=i\tilde{I}[\overline{\xi }_{I}(S,\beta\phi _{b})]$$

If we increase the value of $m$ or $\phi _{b}$, the value of $S_{crit}$
decreases to zero and eventually the branch associated with the Euclidean
regime vanishes.
Furthermore we can see that $S^{I}_{crit}(cl)$ as a function of the
anisotropy $\beta$, becomes infinite at the points of maximum anisotropy. See figure $3$.

\section{Steepest Descent Contour}

After studying the analytical and asymptotic properties of the action
we can now focus on the Euclidean path integral that yields the wave
function of the Universe.

\be
\label{ppsi2}
\Psi [  S,\beta,\phi _{b}]=\int_{C} D\xi  D\phi _{i}e^{-I[\xi, S,\beta ,\phi _{i},\phi _{b}]}   
\ee
In our simplified models the result obtained from a contour $C$ is not
very sensitive to the choice of measure if we stick to the usual measures,
i.e., polynomials of the squared edge lengths. So we shall take

\be
 D\xi D\phi _{i}=\frac{ds_{i}}{2\pi il^{2}}d\phi _{i}=\frac{S}{2\pi iH^{2}}d\xi d\phi _{i}  
\ee

As we have mentioned above there is as yet no universally accepted
prescription for the integration contour to use in quantum
cosmology. Following Hartle \cite{h2}, we shall accept that the main
criteria any contour should satisfy are that it should lead to a
convergent path integral and to a wave function predicting classical
Lorentzian spacetime in the late Universe. The steepest descents
contour over complex metrics seems to be the leading candidate.   
In the simplicial framework, complex metrics arise from complex-valued
squared edge lengths, $(\ref{simpmet})$. The boundary squared edge
lengths, $S$, and $\beta$ have to be real and positive for obvious
physical reasons. But the interior squared edge length, $\xi $, can be
allowed to take complex values.

In general, a SD contour associated with an extremum ends up either at
$\infty $, at a singular point of the integrand, or at another
extremum with the same value of $Im(I)$. We have seen that  when $S$ is big enough the only classical
solutions are a pair of real Lorentzian solutions $(\overline{\xi
}_{I}(S,\beta,\phi _{b}),\overline{\phi }_{i}(S,\beta,\phi _{b}))$ and
$(\overline{\xi }_{II}(S,\beta,\phi _{b}),\overline{\phi }_{i}(S,\beta,\phi
_{b}))$, where  $\overline{\xi }_{I}=\overline{\xi }_{II}<min(1/4,\beta/4)$. They
are located on the first and second sheets respectively, and so have
pure imaginary actions of opposite sign. Given that their actions are
different valued no single SD path can go directly from one to the other
extremum. On the other hand given that

$$ I[\overline{\xi}]=[I[\overline{\xi}^{*}]]^{*} $$

and $$ I[\overline{\xi}_{I}]=- I[\overline{\xi}_{II}]$$
where $*$ denotes complex conjugation, we see that the SD path that passes through  $\overline{\xi }_{II}$ will be
the complex conjugate of the SD path that passes through $\overline{\xi }_{I}$.
So the total SD contour will always be composed of two complex
conjugate sections, each passing through one extremum, and this
together with the real analyticity of the action guarantees that the
resulting wavefunction is real.

The  SD contour passing
through the classical solution
$\{\xi_{cl}(S,\beta,\phi_{b}),\phi_{i}^{cl}(S,\beta,\phi_{b})\}$ is

\be
C_{SD}(S,\beta,\phi _{b})=\biggl\{(\xi \in R,\phi_{i}) : Im[I(S,\xi ,\beta,\phi
_{i},\phi _{b})]=\tilde{I}[\xi_{cl}(S,\beta,\phi_{b}),\phi_{i}^{cl}(S,\beta,\phi_{b})]  \biggr\}
\ee
where $R$ is the Riemann sheet of the action, and $\tilde{I}(\xi)=iI(\xi)$.

In figure $4 $ we show the result of a numerical computation of this
contour for $m=1$, $\phi_{b}=1$, $\beta=1.5$ and $S=50$.

The behaviour is similar to that for other values of the above
variables. Going upward from the extremum, the SD contour proceeds to
infinity in the first quadrant along the curve

\begin{eqnarray*}
 & &\frac{5}{8}\sqrt{(\beta+1)(-\beta^{2}+3\beta-1)}\biggl[1+\frac{1}{60}\frac{m^{2}l^{2}}{H^{2}}
(4\phi_{b}^{2}l^{2}+\phi_{i}^{2}l^{2})\biggr]  \\                         
 &\times & \mbox{}   \frac{S}{H^{2}}(S-S_{crit}^{I})Im(\sqrt{\xi})=\tilde{I}[\xi_{cl},\phi_{cl}]
\end{eqnarray*}
The convergence of the integral along this part of the contour for any
polynomial measure is
guaranteed by the asymptotic behaviour of the action on the first sheet

\begin{eqnarray*}
Re [I^{I}(\xi,\beta ,S,\phi _{i},\phi _{b})]&\sim &
\frac{5}{8}\sqrt{(\beta+1)(-\beta^{2}+3\beta-1)}
\biggl[1+\frac{1}{60}\frac{m^{2}l^{2}}{H^{2}}(4\phi_{b}^{2}l^{2}+\phi_{i}^{2}l^{2})\biggr]\\
                                            &\times &\mbox{}\frac{S}{H^{2}}(S-S_{crit}^{I})\sqrt{\mid\xi\mid}
\end{eqnarray*}

As we move downward from the classical solution we immediately
cross the branch cut and the contour enters the second sheet. The
contour then proceeds to cross the branch cut once more, this time
between $1/4$ and $\beta/4$, emerging onto the third sheet where it
finally proceeds to infinity inside the first quadrant along the
curve

\begin{eqnarray*}
 & &\frac{5}{8}\sqrt{(\beta+1)(-\beta^{2}+3\beta-1)}\biggl[1+\frac{1}{60}\frac{m^{2}l^{2}}{H^{2}}
(4\phi_{b}^{2}l^{2}+\phi_{i}^{2}l^{2})\biggr]  \\                         
 &\times & \mbox{}   \frac{S}{H^{2}}(S+S_{crit}^{III})Im(\sqrt{\xi})=\tilde{I}[\xi_{cl},\phi_{cl}]
\end{eqnarray*}

As on the first sheet the convergence of the path integral along
this section of the contour is guaranteed by the asymptotic behaviour
of the real part of the Euclidean action along it

\begin{eqnarray*}
Re [I^{III}(\xi,\beta ,S,\phi _{i},\phi _{b})]&\sim &
\frac{5}{8}\sqrt{(\beta+1)(-\beta^{2}+3\beta-1)}
\biggl[1+\frac{1}{60}\frac{m^{2}l^{2}}{H^{2}}(4\phi_{b}^{2}l^{2}+\phi_{i}^{2}l^{2})\biggr]\\
                                            &\times &\mbox{}\frac{S}{H^{2}}(S+S_{crit}^{III})\sqrt{\mid\xi\mid}
\end{eqnarray*}

\subsection{Semiclassical Approximation}

From our study of the SD contour for $S>S_{crit}$, we can conclude that the range of
integration in the neighbourhoods of the classical Lorentzian solutions
 give the dominant contribution to the integral, since the
contribution of the other critical points, i.e., the infinities, is
negligible given the asymptotic behaviour of the action. So in order
to obtain the relevant information about the wave function of the
Universe it is not
really necessary to do the full computation of the integral along the
SD contour. A semiclassical approximation based on the classical
solutions found above will suffice. 

Since for $S>S_{crit}$  these  are real Lorentzian
 solutions  with purely imaginary actions
$I_{k}=i\tilde{I}[\xi_{k}^{cl}(S,\beta,\phi
 _{b}),\phi_{i}^{cl}(S,\beta,\phi _{b})]$, then for $S>S_{crit}$, the
 semiclassical approximation is

\begin{eqnarray*}
\Psi _{SC}(S,\phi _{b})&\sim &  \sum_{k=I,II}\sqrt{\frac{S^{2}}{2\pi
H^{4}det\{\frac{\partial^{2}\tilde{I}}{\partial
x_{i}\partial x_{j}}\} }}e^{-i[
\tilde{I}_{k}(S,\beta,\phi _{b})  -\frac{\pi}{4}]}\\
                       &\sim & \mbox{}
\sqrt{\frac{S^{2}}{2\pi H^{4}det\{\frac{\partial^{2}\tilde{I}}{\partial
x_{i}\partial x_{j}}\}}}2\cos{\biggl[\tilde {I}_{cl}(S,\beta ,\phi_{b})-\frac{\pi}{4}\biggr]}
\label{prefac}
\end{eqnarray*}

This semiclassical approximation is specially good when the integrand
is sharply peaked about the classical solutions (extrema), which is
particularly true when the argument of the exponential is large. This
will be the case for the large $S$ of the late Universe and for the
whole range of $S$ when $H^{2}=\Lambda l^{2}/3$ is sufficiently small
as it is the case of our late Universe.

In figure $5$, we show the result of the numerical computation of one
such semiclassical approximation. It is clear that the wave function
exhibits the oscillatory behaviour that characterises the prediction
of classical Lorentzian spacetime in the late Universe as desired.

For $S<S_{crit}$ the situation is not so simple as there are two pairs
of classical Lorentzian solutions as well as the usual pair of
Euclidean solutions. A semiclassical approximation could conceivably
be based on any of these pairs. Furthermore, as we increase the
values of $\phi_{b}$ and $m$, the value of $S_{crit}$ sharply
decreases and so we can envisage a situation where the range of existence
of these solutions practically vanishes. 

Computing the wave funtion for $S=S_{crit}^{I}$, ( figure $6$), we see
that as in the pure gravity case , \cite{fur}, the wave function 
 peaks for universes with large anisotropy, and small  values
of the scalar field. If as in the continuum case
we consider a scenario of quantum nucleation of the Lorentzian universe
at $S=S_{crit}^{I}$, this result seems to favour the universes with
larger anisotropy.

\section{Conclusions}

We have found that the results obtained by us in \cite{meu} can be extended
to anisotropic models. In particular, following \cite{fur}, we have considered
the simplicial minisuperspace based on the cone over the simplest
anisotropic triangulation of the three-sphere, coupled to a massive
scalar field, $\phi_{k}$. The anisotropy is
reflected in the existence of two different kinds of boundary edge
lengths. However, since we admit only one kind of internal edge
length,   the minisuperspace is still two-dimensional as in the
isotropic models considered in \cite{meu}. This means that the path
integral will not involve any new variables. We have found that for
the late Universe the only classical solutions are pairs of real
Lorentzian spacetimes like our own. We showed not only that there is a steepest
descents contour going through these classical solutions but also that
it yields a convergent path integral. These SD contours are very
similar to those obtained in \cite{meu}, but their behaviour is slightly
more complex due to the existence of a larger number of branch points
in the action function. As in the isotropic models, the semiclassical
approximation is quite good, specially when $S$ is large or when $H$ is
sufficiently small, as it is in our late Universe. The computation of
the semiclassical wavefunction shows its oscillatory behaviour,
characteristic of the prediction of   classical Lorentzian spacetimes,
as desired.

Combined with the results in \cite{meu} and \cite{fur} , the results of
this paper show not only the versatility of the simplicial approach to
minisuperspace models, but also that their predictions are generically
in agreement with the results from similar continuum models. In all of
them we find that by making the natural choice of the integration
contour as the SD contour, the wave functions all predict classical
Lorentzian spacetimes for the late Universe.

\vspace{12pt}

{\bf Acknowledgements}

 The work of C. Correia da Silva was supported by the Portuguese
 Ministry of Science and Technology under grant PRAXIS XXI BD-5905/95.
Other support has come from the UK Particle Physics and Astronomy
 Research Council.

\end{document}